\documentclass[aps,showpacs,showkeys,nofootinbib]{revtex4}
\usepackage{amsmath}
\usepackage{graphicx}
\usepackage{dcolumn}
\usepackage{bm}
\usepackage{amssymb}
\usepackage{latexsym}
\usepackage{enumitem}
\usepackage{slashed}
\usepackage{amsfonts} 
\usepackage{color}
\usepackage{lineno}
\usepackage{soul}
\usepackage{amssymb}
\usepackage{mathrsfs}
\usepackage{braket}
\begin{document}

\title {The Polyakov loop dependence of bulk viscosity of QCD matter}

\author{Debmalya Mukhopadhay$^{a}$\footnote{Corresponding author.}}
\email{debphys.qft@gmail.com}
\author{Jan-e Alam$^{a}$}
\email{jane@vecc.gov.in}
\author{R. Kumar$^{b}$}
\email{raviphynuc@gmail.com}
\affiliation{{\it $^{a}$Variable Energy Cyclotron Centre, 1/AF, Bidhan Nagar, Kolkata--700064, India}\\
{\it $^{b}$Department of Physics \& Astrophysics, University of Delhi, New Delhi--110007, India}}

\begin{abstract}
In this work, we show the dependence of bulk viscosity on Polyakov loop in 3+1 dimensional topologically massive model (TMM). 
This model contains equally massive non-Abelian gauge fields without spontaneous symmetry breaking. In earlier works, the 
bulk viscosity was found from the trace anomaly in massless $\phi^4$ model and Yang-Mills (YM) theory and its dependence on 
the quantum corrections was established. In TMM, the trace anomaly is absent due to the presence of kinetic term of a 
two-form field $B$ in the action. This model also provides the dependence of bulk viscosity on the mass of the gauge bosons. 
The mass of the gauge bosons in TMM acts as magnetic mass in the perturbative thermal field theory. This magnetic mass is 
gauge independent unlike what is found in massless YM theory. We also observe that the strong coupling constant has the same
behaviour at high energy limit (i.e. asymptotic freedom) as that of massless YM theory at zero temperature.
\end{abstract}



\maketitle

\section{Introduction}
\label{Section1}
In recent times, Schwarz-type topological field theory in 3+1 dimensions drew huge attentions due to  some of its very 
important characteristics in gauge theory~\cite{momen,mann,jerzy}. One of the interesting features of the model is that 
it contains the massive vector modes in spite of unbroken global symmetry i.e. without taking any recourse of  
Higgs mechanism~\cite{cs:1974,aab:1990}. 
TMM in 3+1 dimensions  carries many advantages in the perturbative analysis in both zero and finite temperature field 
theories over the massless YM theory. The interest is increased manifold  when the model was found to be 
unitary~\cite{meig:1983, al:1997} and renormalizable~\cite{al:01}. In this model,  YM fields become equally massive 
without leaving any extra degrees of freedom unlike the case of Higgs mechanism. Their masses are generated due to the 
presence of  topological term $m B\wedge F$ which contains a quadratic mixing of a one-form YM field $A$ and a 
two-form field $B$.  The massive one- and two-form fields have the same number of degrees of freedom~\cite{kr:1973,dewitt}. 
Hence, an effective theory, constructed by integrating out YM field or $B$ field, becomes a massive theory of 
vector bosons. Unlike the massless YM theory, the complete propagator of YM field in TMM carries a non-zero pole, 
which is the coefficient of topological term $B\wedge F$ in the model. In TMM, the YM field acquires an additional 
physical longitudinal mode due to its mass. But, in spite of having longitudinal mode,  the high energy behavior of 
scattering matrix maintains unitarity in the scattering process involving those modes. This is because of the fact 
that the model is invariant under Becchi-Rouet-Stora-Tyutin (BRST) symmetry transformations~\cite{meig:1983,al:1997,rohit1,rohit2}. 
The unitarity is also maintained at every order of quantum corrections since it is renormalizable~\cite{al:01}. 
The massiveness of the gluon also plays a crucial role in maintaining the cluster decomposition principle i.e. 
causality~\cite{stro, haag} in quantum field theory. On the other hand,  the non-zero pole of gluon propagator can 
explain gluon confinement in  quantum chromodynamics (QCD)~\cite{kugo,fischer,chaichian}. In the regime of strong 
interaction, we shall see that the TMM provides the same asymptotic behaviour of strong coupling at high energy limit 
(i.e. asymptotic freedom) as found in massless YM theory~\cite{gross,politzer,sc:1973}. The asymptotic 
freedom is a 
very significant characteristic of strong sector in the Standard Model.   Beside these important advantages of massive 
YM field at zero temperature, we consider its significant role in the perturbative thermal field theory (TFT). The 
masses of the vector fields put an infrared (IR) cut-off in the model, which behaves as magnetic mass and overcome 
the Linde infrared problem in TFT~\cite{linde, furusawa}. This assures the validity of perturbative analysis of 
the dynamics of YM field at finite temperature which is absent in massless case~\cite{furusawa}. Moreover, it is 
interesting to point out that the magnetic mass in massless YM theory is not gauge independent~\cite{kapusta, bellac}, 
hence it is not a physical quantity. But the lattice gauge theories have showed the short range behaviour of 
chromomagnetic field~\cite{billoire,degrand} due to the presence of physical magnetic mass. In TMM, the mass of the 
gluon is gauge independent~\cite{aab:1990} and it plays the role of magnetic mass in the perturbative regime. It was 
also observed that massless gluons make the QCD vacuum unstable in the formation of bound state~\cite{fukuda,huang}. 
This problem can be cured in this model (without breaking global $SU(N)$ symmetry) because of the massiveness of gluons. 
These essential characteristics motivate us to consider the TMM at finite temperature. We have already established 
the hard thermal loop effective action for TMM in~\cite{debmalya}.

We now consider the transport phenomenon in topologically massive gluonic fluid.  This can be considered in the analysis 
of quark gluon plasma~(QGP)  formed in relativistic heavy ion collision. 
The QGP produced around mid-rapidity at top RHIC (Relativistic Heavy Ion Collider) 
and LHC (Large Hadron Collider) energies 
will contain a negligibly small number of net baryon ({\it i.e.} (number of
quarks) - (number of antiquarks) $\approx 0$) (see Ref.~\cite{Busza}
for a review). The bulk 
thermodynamic properties of such  systems can be
described by a single thermodynamic variable, temperature ($T$)
as the corresponding baryonic chemical potential is negligibly small
due to negligible small number of net baryons (quarks in this case).
Therefore, the result of the present work can be applied to the system formed
at top RHIC and LHC energies.  
Moreover, the tiny gluonic mass present in TMM may not cause 
a severe problem for such an application. 
However, the system formed at lower RHIC energies, at
the upcoming Compressed Baryonic Matter (CBM) experiment at
Facility for Anti-proton and Ion Research(FAIR) and
{\bf{N}}uclotron based {\bf{I}}on {\bf{C}}ollider 
f{\bf{A}}cility (NICA)
at the Joint Institute for Nuclear Research (JINR) will 
have a large net baryons ({\it i.e.} the number of quarks-antiquarks is large
at mid-rapidity).
The system formed  in these collisions will have a 
large baryonic chemical potential~\cite{CBM} and hence,
both temperature and baryonic chemical potential are required to
describe such systems. Therefore,  the results of
the present work can not be applied to such systems. 

QGP is created in a state slightly 
away from equilibrium characterised by various transport coefficients in TFT. The shear and  bulk viscous coefficients 
of a fluid are useful quantities to characterize it. These quantities are required as inputs to solve the relativistic
viscous hydrodynamical equations which have been used in the description of the space-time evolution  of the strongly 
interacting QCD matter formed in nuclear collisions at relativistic energies. To understand the properties of QCD matter, 
it is important to reliably estimate both the shear and bulk viscous coefficients. Like shear viscosity~\cite{jeon,moor1,moor2}, bulk viscosity also carries 
crucial physical significance which are discussed in many recent 
works~\cite{bernhard, alford, bemfica, czajka, astrakhantsev, hattori, Ryu}. We consider the bulk viscosity in pure 
topologically massive gluondynamics. The lattice simulation finds a non-zero bulk viscosity~\cite{meyer} 
in pure gluondynamics.  
To make the analysis simpler, the linear response theory (LRT)~\cite{bala1} is taken into consideration
in the present work. It helps to get the 
coefficients from the linear perturbation around equilibrium state of the fluid. As a consequence, the bulk viscosity $\zeta_T$ can 
be calculated from the well-known Kubo formula~\cite{jeon,moor1,moor2,benincasa}
\begin{eqnarray}
\zeta_T(\omega)&=&\frac{1}{18}\lim_{\omega\to 0}\frac{1}{\omega}\int_{-\infty}^{\infty} dt~e^{-i\omega t} \nonumber\\
&&\times \int d^3x \Braket{\left[\Theta^i_i(x,t), \,\Theta^i_i(0,0)\right]},
\label{em-emcoreltn}
\end{eqnarray}
where $\Theta^{\mu}_{\nu}$ is the energy momentum tensor density of field theory under consideration and $\omega$ is the 
frequency appearing through the Fourier transformation of the correlation  between the spatial trace of the energy 
momentum tensor (EMT) densities.  We observe in Eq.~(\ref{em-emcoreltn}) how the bulk viscosity $\zeta_T$ depends on 
the trace of EMT densities of the quantum fields. Here, the Kubo formula is obtained from the linear response theory 
(LRT)~\cite{kapusta, jeon1, bala1, bala2} which reflects the assumption that the system maintains local equilibrium. 
Since, energy  $\int d^3x~ \Theta^{00}(x)$ of the system is conserved,  we  can shift the spatial trace $\Theta^i_i$ 
by the energy or any multiple of the energy. 
The bulk viscosity diverges near the critical point. 
The diverging nature can be taken into account
phenomenologically by expressing $\zeta_T$ in terms of the correlation length ($\xi$) as discussed below
through Eq.~\ref{zeta_cr}.

The bulk viscosity was found in massless models like $\phi^4$ and YM theories~\cite{moor1,moor2,benincasa} where 
the conformal symmetry is obeyed classically~\cite{dowker}. A common characteristic of these models is that the 
classical conformal invariance breaks down due to the quantum correction in renormalization procedure (i.e. the spectral 
function corresponding to the  correlation of energy momentum tensors depends on the breaking of conformal symmetry). 
The exceptional case is found in $\mathcal{N}=4$ super YM theory where it is observed that $\zeta_T=0$~\cite{benincasa}. 
The trace anomaly causes problems when we consider the theories in curved spacetime.  It was found that this anomaly 
causes violation of the fifth axiom in the construction of its uniqueness in curved spacetime which affects significantly 
in the semi-classical treatment of general theory of relativity~\cite{wald, stelle}. This carries a great importance 
in the consideration of de-Sitter spacetime (i.e. maximally symmetric spacetime with positive cosmological constant 
$\Lambda>0$)  where the trace of the EMT  can determine the structure of full EMT.  On the other hand, 
trace anomaly provides negative vacuum energy density from the perturbative regime~\cite{shifman}  which contradicts 
with the cosmological observations~\cite{ralf}.  The bulk viscosity may also lead to an alternative to the dark energy 
in cosmological scenario in de-Sitter spacetime~\cite{sebastein}.

From Eq.~(\ref{em-emcoreltn}), we observe that the  bulk viscosity depends on the correlation of energy momentum densities. 
We get the correlation from the low energy theorem (LET) at finite temperature~\cite{ellis, sushpanov, karsch} as:
\begin{widetext}
\begin{eqnarray}
\left(T\frac{\partial}{\partial T}-4\right)^n\braket{\Theta^\mu_\mu}
= \int d\tau_n d^3x_n\cdots d\tau_1 d^3x_1 \Braket{\Theta^\mu_\mu(\tau_n, x_n)\cdots\Theta^\mu_\mu(\tau_1, x_1)
\Theta^\mu_\mu(0, 0)},
\label{let}
\end{eqnarray}
\end{widetext}
where gulons degrees of freedom are relevant. We take background field  method for the calculation of  l.h.s. of  
Eq.~(\ref{let}). This calculation provides the vacuum expectation value of regularized trace of EMT density at 
finite temperature. With this purpose, we will construct the one-loop effective action using heat kernel 
method~\cite{dewitt, christensen1, christensen2, barvinsky, filho, Vassilevich} at finite temperature. In this 
procedure, it was already observed~\cite{megias1, megias2} that the heat kernel coefficients contain a holonomy: 
$L(x, \beta)= \mathcal{T}\exp\left(-\int_{x_0}^{x_0+\beta} A_0(x'_0, \textbf{x})~ dx'_0\right)$; $L(x, \beta)$ is 
known as untraced Polyakov loop\footnote{Here $A_0$ is the temporal component of quantum gauge field.}~\cite{polyakov} 
and $\mathcal{T}$ represents time-ordered product. It arises in the calculation due to the compactification of 
fourth Euclidean axis in thermal field theory.  Hence, the modified heat kernel method, found in~\cite{megias1,megias2}, 
includes the contribution of  $L(x, \beta)$ in the construction of effective action. This causes the dependence 
of $\zeta_T$ on $L$ and we shall show it in the next section. Specifically, we consider the behavior of spectral function 
\begin{eqnarray}
\rho(\omega)=\int_{-\infty}^{\infty} dt~e^{-i\omega t}\int d^3x \Braket{\left[\Theta^\mu_\mu(x,t), \Theta^\nu_\nu(0,0)\right]}, 
\label{bulkvisdiagm1}
\end{eqnarray}
from the model. In the massless YM theory, the bulk  viscosity $\zeta_T$ is calculated from quantum corrections 
of the trace of EMT density of the YM field (i.e. the leading contribution in the calculation comes from the 
conformal or trace anomaly).  This causes the dependence of $\zeta_T$  on the strong coupling in the perturbative 
computation. In the lattice QCD~\cite{meyer}, the ratio $\frac{\zeta_T}{s}$ was computed for pure gluondynamics 
where $s$ is the entropy density. But, in the case of TMM, we shall observe how the leading order in the spectral 
function depends on the mass of gauge fields and expectation value of untraced Polyakov loop. The present 
investigation emphasizes the possibility of finding bulk viscosity in the perturbative regime of QCD with the 
same asymptotic freedom as found in the literature.

In this endeavor, we present an explicit calculation of the spectral function in Sec. \ref{Section2}. 
Section \ref{Section3} contains the discussion, conclusions and the future aspects of the model in the realm of
thermal field theory. We take the signature of the Minkowski metric $\eta_{\mu\nu}$ as $(+,~-, ~-, ~-)$. We 
have also taken the convention: $\hbar=k_B=c=1$ where $k_B$ is Boltzmann's constant.  

\section{Calculation}
\label{Section2}
Within the scope of LRT, the hydrodynamical transport coefficients by using Green-Kubo formula can be written as follows:
\begin{widetext}
\begin{eqnarray}
\eta(\omega)\left( \delta_{k\{i}\, \delta_{j\}l} - \frac{2}{3}\delta_{kl} \, \delta_{ij}\right)
+ \zeta_T(\omega)\delta_{ik}\, \delta_{lm} 
=\lim_{\omega\to 0} \frac{1}{\omega}\int d^3x \int_0^\infty dt~ e^{i(\omega t
- \textbf{k}\cdot \textbf{x})} \Braket{\left[\Theta_{ij}(t, \textbf{x}),\, \Theta_{kl}\right]}, 
\end{eqnarray}
\end{widetext}
where $\eta(\omega)$ is called as  shear viscosity and $\delta_{m\{a}\delta_{b\}n}=\dfrac{1}{2}\left(\delta_{am}\delta_{bn}+\delta_{bm}\delta_{an}\right)$. The bulk viscosity can be obtained from the above formula 
by contracting $i$, $j$ and $k$, $l$ as 
\begin{eqnarray}
\zeta_T(\omega)=\lim_{\omega\to 0} \frac{1}{9\omega}\int d^3x \int_0^\infty dt~ e^{i(\omega t-\textbf{k}\cdot \textbf{x})}\Braket{\left[\Theta^j_{j}(t, \textbf{x}), \, \Theta^k_{k}\right]}. \nonumber\\
\end{eqnarray}
 The energy momentum tensor is obtained from the following part of the action in Minskowski spacetime 
\begin{eqnarray}
S_0 &=&\int d^4x \Big(-\frac{1}{4}F^{a\mu\nu} F^a_{\mu\nu}+\frac{1}{12}H^{a\mu\nu\lambda} H^a_{\mu\nu\lambda} \nonumber\\
&& \qquad \qquad \qquad + \frac{m}{4}\epsilon^{\mu\nu\alpha\beta} B^a_{\mu\nu} F^a_{\alpha\beta}\Big),
\label{topactn}
\end{eqnarray}
where $F^a_{\mu\nu}=\partial_{\mu}A^a_{\nu}-\partial_{\nu}A^a_{\mu}+ g f^{abc} A_\mu^b A_\nu^c$ and $H^a_{\mu\nu\lambda} =  \partial_{[\mu} \, B^a_{\nu\lambda]} + g f^{abc}\, A^b_{[\mu} \, B^c_{\nu\lambda]} - g  f^{abc} \,F_{[\mu\nu}^b\, C_{\lambda]}^c$ are the field strength of YM field and tensor field, respectively,  and $f^{abc}$ is the structure constant of $SU(N)$ group. The presence of an auxiliary field $C^a_\mu$ in the expression  of $H^a_{\mu\nu\lambda}$ assures the invariance of the action under the following gauge transformations:
\begin{eqnarray}
&& A_{\mu}^a \to A_{\mu}^a, \qquad B_{\mu\nu}^a \to B_{\mu\nu}^a + \left(D_{[\mu}\, \theta_{\nu]}\right)^a,\nonumber\\
&&  C_\mu^a \to C_\mu^a + \theta_{\mu}^a,
\end{eqnarray}
where $\theta_{\mu}^a$ is a vector field in the adjoint representation of $SU(N)$.  Including ghosts' sectors, 
we have the full action as given by
\begin{eqnarray}
S = S_0 & + & \int d^4x \Big[h^af^a + \frac{\xi}{2}\, h^a h^a + h_\mu^a \big(f^{a\mu} 
+ \partial^\mu \widetilde{n}^a \big) \nonumber\\
&+& \beta^a \big(D_\mu \beta^a - g f^{abc}\, \omega_{\mu}^b\, \omega^c \big) 
+ \frac{\widetilde{\eta}}{2}\, h_\mu^a \,h^{a\mu} \nonumber\\
&-& \partial_\mu \bar{\omega}^{a\mu}\, \alpha^a + \bar{\alpha}^a\, \partial_\mu \omega^{a\mu} 
+ \widetilde{\zeta} \,\bar{\alpha}^a \,\alpha^a + {\bar \omega}^a\, \partial_\mu D^\mu \omega^a \nonumber\\
&+&  \bar{\omega}^a_\mu \big\{g f^{abc} \partial_\nu \big(B^{b\mu\nu}\,\omega^c \big) 
+ \partial_\nu \big(D^{[\mu}\, \omega^{\nu]}\big)^a \nonumber\\
&+& \partial_\nu \big(g f^{abc}\, F^{b\mu\nu} \, \theta^c)\big \} \Big],
\label{nonabactn}
\end{eqnarray}
where $S_0$ is the action given in Eq.~(\ref{topactn}) and  
$ f^a = \partial^\mu A^a_{\mu}$,  $f^a_\mu = \partial^\nu B^a_{\nu\mu}$. The parameters $\xi, \widetilde{\eta}$ and $\widetilde{\zeta}$ are the dimensionless 
gauge-fixing parameters. The auxiliary fields $h^a$ and $h^a_\mu$ play the role of Nakanishi-Lautrup type fields. 
Here  $(\bar{\omega}^a) \omega^a$ and  $(\bar{\omega}^a_\mu) \omega^a_\mu$ (with ghost number $(-1)+1$) 
are  the Fermionic scalar and vector (anti-)ghost fields for  the vector gauge field $A^a_\mu$ and tensor field 
$B^a_{\mu\nu}$, respectively.  The bosonic scalar fields $(\bar \beta^a) \beta^a$ (with ghost number $(-2)+2$) are the 
(anti-)ghost fields for the Fermionic vector (anti-)ghost fields and $\widetilde{n}^a$ is the bosonic scalar ghost field (with 
ghost number zero). These scalar ghost fields are required for the stage-one reducibility of the two-form field. Furthermore,  
$\alpha^a$ and $\bar{\alpha}^a$ are the Grassmann valued auxiliary fields (with ghost number $+1$ and $-1$, respectively).  
This model contains massive non-Abelian gauge field and it was shown to be BRST 
invariant~\cite{al:1997,rohit1,rohit2}. 
In~\cite{rohit1,rohit2}, it is seen that the model is also invariant under the anti-BRST symmetry transformations. 
The $\mathcal{CP}$ symmetry is not violated in this model.

The EMT density corresponding to the action [cf. Eq.~{(\ref{topactn})}] in curved space time is given by
\begin{eqnarray}
\Theta_{\mu\nu}=\frac{2}{\sqrt{-\widetilde{g}}} \frac{\delta \widetilde{S}_0}{\delta g^{\mu\nu}},
\end{eqnarray} 
where $\widetilde{S}_0=\int\sqrt{-\widetilde{g}}  \Big(-\frac{1}{4}F^{a\mu\nu} F^a_{\mu\nu}+\frac{1}{12}H^{a\mu\nu\lambda} H^a_{\mu\nu\lambda}$ $+\frac{m}{4}\epsilon^{\mu\nu\alpha\beta} B^a_{\mu\nu} F^a_{\alpha\beta}\Big)d^4x$ and  $g_{\mu\nu}$ is the metric in the curved spacetime. Here $\widetilde{g}= \text{Det}g_{\mu\nu}$. We find that the EMT density 
corresponding to the action of TMM classically as
\begin{eqnarray}
\Theta_{\mu\nu}=\Theta^{YM}_{\mu\nu}+\frac{1}{2}\,\left(H^a_{\mu\alpha\beta} H_\nu^{a~\alpha\beta}-\frac{1}{6}\,g_{\mu\nu} H^{a\alpha\beta\rho} H^a_{\alpha\beta\rho}\right), \nonumber\\
\label{emt}
\end{eqnarray}
where  $\Theta^{YM}_{\mu\nu}=-F^a_{\mu\alpha}F^{a~\alpha}_{\nu}+\dfrac{1}{4}\,g_{\mu\nu}F^{a\alpha\beta} F^a_{\alpha\beta}$ is the standard EMT for the YM field $A_\mu^a$.  Since the topological term is invariant under the variation of metric tensor, hence, it does not provide any contribution in TMT in Eq.~(\ref{emt}). The trace of $\Theta^\mu_\nu$ is non-zero in 3+1 dimensional spacetime reads
\begin{eqnarray}
\Theta^{\mu}_\mu= \frac{1}{6}\,H^{a\rho\nu\lambda} H^a_{\rho\nu\lambda},
\label{trace}
\end{eqnarray}
because $ g^{\mu\nu}\Theta^{YM}_{\mu\nu}=0$. Hence, it is clear from Eq.(\ref{trace}) and the action in Eq.~(\ref{topactn}) 
that the kinetic term of $B$ field is responsible for the absence of conformal symmetry at zero temperature.  This implies 
that we can find the bulk viscosity for the topologically massive  YM fluid at finite temperature. For our purpose, we use 
low energy theorem at finite temperature~\cite{ellis,sushpanov,karsch}.
We use the splitting\footnote{The quantum fields are designated by lowercase letters having Lorentz and gauge indices.} 
\begin{eqnarray}
A^e_\mu = \mathcal{A}^e_\mu+a^e_\mu,
\label{split}
\end{eqnarray}
where $\mathcal{A}^e_\mu$ is background and $a^e_\mu$ is the quantum fields.    
Due to the splitting, the YM field strength becomes 
\begin{eqnarray}
F^e_{\mu\nu} ( \mathcal{A}, a)= F^e_{\mu\nu} (\mathcal{A})+\left(D^{\mathcal{A}}_{[\mu}\, a_{\nu]}\right)^e 
+  g \,f^{ecd}\,a_\mu^c \,a_\nu^d,
\end{eqnarray}  
where $e$ is a gauge index. The covariant derivative $D^{\mathcal{A}}_\mu=\partial_\mu+ g \mathcal{A}_\mu$ is taken with respect to the background field 
$A_\mu^a$. Here $g$ is the gauge coupling constant.  Our aim is to find an effective action at one-loop level from Eq.~(\ref{nonabactn}), which can provide regularized EMT density as shown in ~\cite{dewitt, christensen1, christensen2}. For this purpose, it is sufficient to find the terms in the action which contain quantum fields $a_\mu$ quadratically. It is to be noted that the YM and $B$ fields are coupled quadratically due to presence of the topological term $B\wedge F$ in the Lagrangian density\footnote {Due to Eq.~(\ref{split}), the following term $\dfrac{m}{4}\epsilon^{\mu\nu\rho\lambda}B_{\mu\nu} F_{\rho\lambda}$ contributes $\dfrac{m}{4}\epsilon^{\mu\nu\rho\lambda}b_{\mu\nu}\left(F_{\rho\lambda}(\mathcal{A})+2\left(D^{\mathcal{A}}_{\rho}a_{\lambda}\right)\right)$ in the quadratic part under consideration.} in Eq.~(\ref{topactn}). This mixing leads us to construct a matrix from the kinetic terms of YM and $B$ fields and $\dfrac{m}{2} B\wedge F$ term \footnote{The $\Delta$ is appeared by the re-expressing $S_0$ [cf. Eq.~(\ref{topactn})] as $S_0=\displaystyle\int d^4x ~\Phi^{\mathbb{T}} \Delta \Phi$ where $\Phi=\binom{A}{B}$ and $\Phi^{\mathbb{T}}$ is the transpose of $\Phi$.}
\begin{eqnarray}
\Delta= \left(
\begin{array}{cc}
\Delta_A & \Delta_{AB}\\
\Delta_{BA} & \Delta_B
\end{array}
\right).
\end{eqnarray} 
It is a block matrix whose determinant is given by~\cite{dewitt}
\begin{eqnarray}
\text{det}\Delta &=& \left(\text{det}\Delta_B \right)^{-1}\text{det}\left(\Delta_A-\Delta_{AB}\Delta^{-1}_B \Delta_{BA}\right) \nonumber\\
&=&\left(\text{det}\Delta_B \right)^{-1}\text{det} \Delta_A ~ \nonumber\\
&& \times \text{det}\left(1-\left(\Delta_A\right)^{-1}\Delta_{AB}\Delta^{-1}_B \Delta_{BA}\right).
\label{blkmatrix}
\end{eqnarray}
From the complete action of the model, we can clearly notice the presence of trilinear couplings among the ghost and gluon fields (due to the above splitting provides the quadratic terms in all quantum fields). The ghosts' sector consists of Faddeev-Popov (FP) ghosts for $a_\mu$, vector ghosts for $B$ field,  FP ghosts of the vector ghost fields and a scalar ghost. The general structure of the effective action, at $T=0$, in one-loop level is given by~\cite{dewitt, sw}
\begin{eqnarray}
W^1[\mathcal{A}]&\propto & \int\sqrt{-\widetilde{g}}  \, \bigg( n_1\text{Tr}\ln\Delta + n_2\text{Tr}\ln\Delta_{gh} \nonumber\\
&& \qquad\quad + n_3\text{Tr}\ln\Delta_{vecgh} + n_4\text{Tr}\ln\Delta_{ghvecgh} \nonumber\\
&& \qquad\quad + n_5\text{Tr}\ln\Delta_{sclgh}\bigg)d^Dx,
\end{eqnarray}
 where $n_i$'s (with $i =1, 2, 3, 4, 5$) are the numerical factors which appear after integrating out the quantum fields in the partition functional. These numerical factors also depend on the spin-statistics of quantum fields~\cite{dewitt, sw}; $\Delta_{\Phi_i}$'s are appearing  from the action after integrating out the quantum fields from the partition functional~\cite{dewitt, sw} 
 \begin{eqnarray}
 Z[\Phi_{i0}]=\int\prod_{i=1}^n \mathcal{D}\widetilde{\phi_i} \exp\left(\frac{i}{2}\displaystyle\int d^4x\sqrt{-\widetilde{g}} \displaystyle\sum_{i=1}^{n=1}\Phi_i \Delta_{\Phi_i} \Phi_i\right), \nonumber\\
 \label{partitionfuncn}
 \end{eqnarray}
 where $\Phi_{i0}$ is the background field  of $i$-th type field: $\Phi_i=\Phi_{i0}+\widetilde{\phi}_i$; $\widetilde{\phi}_i$ is the quantum part of $\Phi_i$. We have suppressed the spin and gauge indices of the fields in Eq.~(\ref{partitionfuncn}) and $\Delta_{\Phi_i}$'s may be the function of background fields or not. For example, we note from the complete action of TMM that $\Delta$,  $\Delta_{gh}$, $\Delta_{vecgh}$ and $\Delta_{ghvecgh}$ are the functions of  covariant derivative with respect to the background YM field $\mathcal{A}$ but for the scalar ghost $\Delta_{sclgh}$ does not contain any covariant derivative and $\mathcal{A}$. One can also clearly check from the structure of the bulk matrix in Eq.~(\ref{blkmatrix}) that $\Delta$ contains $\Delta_A$, $\Delta_{AB}$, and $\Delta_{BA}$. Here,  $\Delta_{A}$ and $\Delta_{B}$ appear from the kinetic terms of $A$ and $B$ fields whereas other $\Delta$'s comes from $B\wedge F$ term. Following the general rule~\cite{dewitt}, we can write the effective action at $T=0$ as 
\begin{eqnarray}
W^1 &=& -\frac{1}{2}\int\sqrt{-\widetilde{g}}  ~\Big[\text{Tr}\ln\Delta-2\text{Tr}\ln\Delta_{gh} 
+ 2 \text{Tr}\ln\Delta_{vecgh} \nonumber\\
&-& 2\text{Tr}\ln\Delta_{ghvecgh}-\text{Tr}\ln\Delta_{sclgh}\Big]d^Dx.
\label{effactn}
\end{eqnarray}
For the computation of  $\text{Tr}\ln\Delta_{\Phi_i}$, we use the heat kernel method at finite temperature. 
The relation between the heat kernel coefficients at zero and finite temperature is given explicitly~\cite{filho, megias1, megias2}. This calculation is done with covariant gauge-fixing condition $D_\mu^{\mathcal{A}}a^\mu=D_\mu^{\mathcal{A}}b^{\mu\nu}=D_\mu^{\mathcal{A}}\omega^\mu=0$. 
We know from the detail of heat kernel method~\cite{dewitt, barvinsky, Vassilevich} how the trace of 
an operator $\Delta_{\Phi_i}$ is calculated from the coincidence limit of a matrix with matrix element is 
\begin{eqnarray}
\text{Tr}\ln\Delta_{\Phi_i}= \int_0^\infty\frac{1}{\left(4\pi\tau\right)^{D/2}}   H(x, x, \tau)~ \frac{d\tau}{\tau},
\label{heatkrnl1}
\end{eqnarray}
where $H(x, x,  s)=\displaystyle\sum_{n=0}^\infty s^n~\text{Tr} a_n $ is called to be heat kernel corresponding to 
an operator $\Delta_{\Phi_i}$ and $a_n$'s are the heat kernel coefficients in $D$-dimensional spacetime. Here $\text{Tr}f(x,y)=\int\sqrt{-\widetilde{g}} ~ \text{tr}\,f(x,x) d^Dx$  and `tr' denotes the trace over the Lorentz and internal indices~\cite{dewitt, barvinsky, Vassilevich} of function $f(x,y)$.  In the construction of the effective action at one-loop level,  we consider the trace [cf. Eq.~(\ref{effactn})] at finite temperature.  The expression in Eq.~(\ref{heatkrnl1}) is modified at finite temperature as~\cite{filho}
 \begin{eqnarray}
 H_\beta(x, y, \tau)= H(x, y, \tau)\left[1+2 \sum_{n=1}^\infty\kappa_n e^{-\frac{n^2\beta^2}{4\tau}}\right],
 \label{heatkrnl2}
 \end{eqnarray}
where $\beta=\frac{1}{T}$ and $\kappa_n$ signifies the dependence of the expansion on the spin-statistics of the fields. In fact,  $\kappa_n=(-1)^n$ for fermionic field and $\kappa_n =1$ for bosonic field. The relation in Eq.~(\ref{heatkrnl2}) was found to be incomplete~\cite{megias1, megias2}. This incompleteness occurs due to the exclusion of $L(x, \beta)$ as mentioned in the 
Sec. \ref{Section1} . The heat kernel expansion of trace: Tr$\left(e^{-s(-D^\mu D_\mu+X)}\right)=\displaystyle\int \dfrac{ds}{s}\dfrac{1}{(4\pi s)^{D/2}} \displaystyle\sum_{n=0}^\infty s^n\text{Tr}(a^T_n)$ is given with the following heat kernel coefficients (upto mass dimension 4)~\cite{megias1, megias2}
\begin{eqnarray}
a_0^T(x,x)&=&\phi_0(L,s),\\
a_1^T(x,x) &=&-\phi_0(L,s) X,\\
a_2^T(x,x) &=&-\frac{1}{2}\phi_0(L,s)X^2-\frac{1}{3}\bar{\phi}_2(L,s)E_i^2 \nonumber\\
&+& \frac{1}{12}\phi_0(L,s)F_{ij}^2, 
\end{eqnarray}
where $E_i= F_{0i}$, and 
\begin{eqnarray}
\phi_0(L,s) &=&\left[1+2\sum_{n=1}^\infty L^n e^{-\frac{n^2\beta^2}{4s}}\right]\label{phi0}, \\
\phi_n(L) &=& \frac{1}{\beta}\sqrt{4\pi s}\sum_{p_{0r}}s^{n/2}~ Q^n_r e^{sQ_r^2},\\
\bar{\phi}_2 &=&\phi_0+2\phi_2,
\end{eqnarray}
where $Q_r=i p_{0r}-\frac{r}{\beta}\ln L$~\cite{megias1, megias2}. Thus,  $\braket{\Theta_{\mu\nu}}$, derived from the effective action, will depend on $L$ and as a consequence of LET [cf. Eq.~(\ref{let})] we can see the correlation among the EMT densities become dependent on $L$, too. Its further consequence is very interesting which we shall show how $\zeta_T$ depends on $L$.

 According to the method outlined in~\cite{dewitt, barvinsky, Vassilevich}, we need to identify  $X$ in the Laplace-type operator appearing in the kinetic terms of quantum fields: 
  \begin{eqnarray}
  \Delta_A\equiv-\frac{1}{2} \left(-D^\mu_{\mathcal{A}}D^{\mathcal{A}}_\mu +X\right),
  \end{eqnarray}  
  where covariant derivative is expressed as $D_\mu^{\mathcal{A}}=\partial_\mu+\widetilde{\omega}_\mu(\mathcal{A})$  and $\widetilde{\omega}_\mu$ is  the ``connection''~\cite{dewitt, barvinsky, Vassilevich}. Generally, $X$ and $\widetilde{\omega}_\mu$ are matrix valued functions in the non-Abelian gauge theory.  Corresponding to the YM field,  we have~\cite{Vassilevich}
\begin{eqnarray}
\left(\widetilde{\omega}_\mu^{cd}\right)^\rho_\lambda &=& -gf^{ecd} \mathcal{A}^e_\mu\delta^\rho_\lambda,\label{omega}\\
\left(X^{cd}_A\right)_{\rho\lambda}&=&  2gf^{ecd}F^e_{\rho\lambda}(\mathcal{A})\label{Ecoef}.
\end{eqnarray}
The expression of the $\widetilde{\omega}_\mu$'s in Eq.~(\ref{omega}) are same for all the quantum fields 
in Eq.~(\ref{nonabactn}) but $X$'s will be different.  
For example,  we get $X_B$ corresponding to the $b_{\mu\nu}$ and vector ghost fields as: 
\begin{eqnarray}
\left(X_B^{cd}\right)^{\mu\nu}_{\rho\lambda}&=& 2g f^{ecd} F^e_{\alpha\beta}(\mathcal{A})\,\eta^{\alpha[\mu}\delta^{~\nu]}_{[\lambda}\delta_{\rho]}^{~\beta},\\
\left(X_{vecgh}^{cd}\right)_{\mu\nu}&=&-gf^{ecd}F^e_{\mu\nu}(\mathcal{A}).
\end{eqnarray}
For the rest of the ghost fields, $X=0$ which can be read-off from the action in Eq.~(\ref{nonabactn}).  
We also have the following explicit expressions
\begin{eqnarray}
(\Delta_{AB})^{\mu\rho\sigma} &=& i\frac{m}{2} \epsilon^{\alpha\mu\rho\sigma}\overrightarrow {D}_{\alpha}^x, \\
(\Delta_{BA})^{\mu\rho\sigma} &=& i\frac{m}{2} \epsilon^{\beta\mu\rho\sigma}\overleftarrow{D}_{\beta}^y, \\
\left(\Delta_B\right)_{\alpha\beta,~\rho\sigma} &=&-\eta_{\alpha[\rho} \,\eta_{\sigma]\beta}D^\mu D_\mu+\left(X\right)_{\alpha\beta,~\rho\sigma}.
\end{eqnarray}
In the above, $\Delta_B$ can be expressed as 
\begin{eqnarray}
\left(\Delta_B\right)_{\alpha\beta,~\rho\sigma}=- \eta_{\alpha[\rho} \, \eta_{\sigma]\beta}\,\partial^\mu \partial_\mu
+\sigma_{\alpha\beta,~\rho\sigma},
\end{eqnarray}
where $\sigma_{\alpha\beta,~\rho\sigma}=\eta_{\alpha[\rho} \,\eta_{\sigma]\beta}\left(2g\mathcal{A}^\mu\partial_\mu+g\partial_\mu \mathcal{A}^\mu+g^2 \mathcal{A}^\mu \mathcal{A}_\mu\right)+\left(E\right)_{\alpha\beta,~\rho\sigma}$. 
We can safely neglect the contribution from $\Delta_{sclgh}$ in the effective action [cf. Eq.~(\ref{effactn})] because of the absence of the background field in  kinetic term of 
the scalar ghost field of $b_{\mu\nu}$. Since, in  the  leading order 
\begin{widetext}
\begin{eqnarray}
(\Delta_{AB} \Delta^{-1}_B \Delta_{BA})^{\mu\nu}_{xy} &=& m^2\delta^4(x-y)\eta^{\mu\nu}\,\left(1-\frac{1}{D}\right) 
+\mathcal{J}^{\mu\nu}_{xy}(g^n,\mathcal{A}), \quad n\geq 1,
\label{leadinodr}
\end{eqnarray}
\end{widetext}
we can re-express $\text{Tr}\ln\Delta$ as 
\begin{eqnarray}
\text{Tr}\ln\Delta &=& \text{Tr}\ln \widetilde{\Delta}-\text{Tr}\ln\Delta_B \nonumber\\
&& \quad + \,\text{Tr}\ln\left(1-\mathcal{J}(g^n,\mathcal{A})\widetilde{\Delta}^{-1}\right),
\label{reexpress}
\end{eqnarray}
 where $\widetilde{\Delta}= \Delta_A +\widetilde{m}^2$ and $\widetilde{m}^2=\left(1-\frac{1}{D}\right)m^2$.   In Eq.~(\ref{leadinodr}), the matrix-valued operator $\mathcal{J}^{\mu\nu}_{xy}(g^n,\mathcal{A})$ designates the parts of $\left(\Delta_{BA} \Delta^{-1}_B \Delta_{AB}\right)^{\mu\nu}_{xy}$ which contains various non-zero powers of $g$ and background YM field $\mathcal{A}$. The significance of the r.h.s. of Eq.~(\ref{reexpress}) will be shown later in our analysis. Hence, we have now
 \begin{eqnarray}
W^1 &=& - \frac{1}{2}\int~\sqrt{-\widetilde{g}}\Big[\text{Tr}\ln\widetilde{\Delta}-\text{Tr}\ln\Delta_B-2\text{Tr}\ln\Delta_{gh} \nonumber\\
&+& \text{Tr}\ln\left(1-\mathcal{J}(g^n,\mathcal{A})\widetilde{\Delta}^{-1}\right)\Big]d^Dx.
\label{modeffectiveactn}
\end{eqnarray}
The last term is found from the series expansion 
\begin{eqnarray}
\ln(1-Y)=-\sum_{n=1}^\infty \frac{Y^n}{n}.
\label{logexpnsn}
\end{eqnarray}
It can be readily checked that the first term in the above expansion is: $- Y = \left(\mathcal{J}(g^n,\mathcal{A})\widetilde{\Delta}^{-1}\right)$. The traces in the first three terms can be found from Eq.~(\ref{heatkrnl1}). Now we explain the significance of Eq.~(\ref{reexpress}). Following Eq.~(\ref{heatkrnl1}), we can write  
\begin{eqnarray}
\braket{x|\ln\Delta|x}=\displaystyle\int_0^\infty \frac{ds}{s}~e^{-\widetilde{m}^2 s}~ \widetilde{H}(x, x, s),
\label{heatkrnel3}
\end{eqnarray}
which shows the trace is convergent in the large-$s$ region. Further, $\widetilde{H}(x, x, s)$ is different from $H(x, x, s)$ [cf. Eq.~(\ref{heatkrnl1})] due to the rearrangement of the terms as shown in Eq.~(\ref{leadinodr}). To find the trace, we should note that the Laplace-type operator acts on the gluon and vector ghost fields in $(N^2-1)D$ dimensional internal space whereas it acts on the FP ghost fields in $(N^2-1)$ dimensional internal space. For the $b_{\mu\nu}$ field, the dimension of the internal space becomes $\dfrac{D(D-1)}{2}(N^2-1)$ where the operator acts. This leads to the following expression for the heat kernel expansion at finite temperature~\cite{megias1, megias2}:
  \begin{eqnarray}
  W^1=-\frac{1}{2}\int_0^{\infty}\frac{ds}{s} \frac{e^{-\widetilde{m}^2s}}{(4\pi s)^{D/2}}\sum_{n=0}^{\infty} \text{Tr}~ a^T_n(x,x) s^n,
  \end{eqnarray}
 where
\begin{eqnarray}
a^T_0 &=& \frac{7D-D^2-8}{2}\phi_0(L), \label{htkrnlceff0}\\
a^T_2 &=& \left[(2-D)+\frac{7D-D^2-8}{24}\right]\phi_0(L) F^{a\mu\nu} F_{\mu\nu}^b N^{ab} \nonumber\\
&+&\frac{7D-D^2-8}{12} E_i^a E_i^b N^{ab} \bar{\phi}_2(L),
\label{htkrnlcoeff2}
\end{eqnarray}  
and 
\begin{eqnarray}
\phi_n(L)= \frac{1}{\beta}\sqrt{4\pi s}\sum_{p_{0r}}s^{n/2}~ Q^n_r e^{sQ_r^2}, \qquad \bar{\phi}_2=\phi_0+2\phi_2, \qquad
\end{eqnarray}
with $Q_r=i\left( p_{0r}-\frac{r}{\beta}\ln L\right)$~\cite{megias1, megias2} and $N^{ab}=f^{acd}\,f^{bcd}= N\delta^{ab}$ in Eq.~(\ref{htkrnlcoeff2}).  Now we are going to get an explicit expression of the effective action at finite temperature using Eq.~(\ref{heatkrnl2}). From the general expression of heat kernel coefficients in~\cite 
{megias1,megias2}, we can write the effective action for massless gluon field at finite  temperature as\footnote{We have suppressed the terms involving the Riemann curvature, Ricci tensor and scalar and their derivatives because they do not contribute in the limit $g_{\mu\nu}\to\eta_{\mu\nu}$.}~\cite{megias2}
\begin{eqnarray}
W^1 &=& \int d^Dx \sqrt{-\widetilde{g}}~\bigg(-\frac{\pi^2}{45}T^4(N^2-1) \nonumber\\
&+& \frac{2\pi^2}{3}T^4~\text{tr}\left[\nu^2(1-\nu)^2\right]+\mathcal{I}(\mathcal{A}, T)\bigg),\quad 0<\nu <1, \nonumber\\
\label{effactnmasslessYM}
\end{eqnarray}
where $\nu=\left(\dfrac{\ln L}{2\pi i}\right)$ and $\mathcal{I}(\mathcal{A}, T)$ designate the terms depending on  both temperature and background YM field. In the massless case, it is already seen in~\cite{megias2} that IR divergence exists in the large-$s$ region. This causes a serious problem in perturbative TFT which we have pointed out in the introduction of the paper. In the TMM, the problem is settled due to the presence of the factor $e^{-\widetilde{m}^2s}$ [cf. Eq.~(\ref{heatkrnel3})]. We need the expression of $\phi_0$ [cf. Eq.~(\ref{phi0})]  
to calculate the dimensionally regularized effective action as~\cite{megias1, megias2}
\begin{eqnarray}
W^{1}=-\frac{1}{2}\int_0^{\infty}\frac{ds}{s} \frac{\mu^{2\epsilon}}{(4\pi s)^{D/2}}\sum_{n=0}^{\infty}\ \text{Tr}~ a^T_n(x,x)~ s^n,
\end{eqnarray}
where $\mu$ is called to be subtraction point and the regularization will be done in $D=4-2\epsilon$  dimensions. To calculate the terms in the leading order of heat kernel expansion, we need to work out the integration of the type~\cite{megias1, megias2}
\begin{eqnarray}
 I_{l, n}=\int_0^{\infty}\frac{ds}{s}~(4\pi\mu^2 s)^{\epsilon}~  s^l~ \phi_n(\omega,s)~ e^{-\widetilde{m}^2s}, \quad |\omega|=1, \quad
 \end{eqnarray}
which yields in the leading order for $l=-2$ and $n=0$
\begin{eqnarray}
I_{-2, 0}&=& \int_0^{\infty}\frac{ds}{s}~ e^{-\widetilde{m}^2s}~(4\pi\mu^2 s)^{\epsilon}  s^{-2} \phi_0(\omega,s)\nonumber\\ 
&=&\nonumber(4\pi\mu^2)^\epsilon \int_0^\infty  ds s^{-2+\epsilon}e^{-\widetilde{m}^2s}\sum_{k\in\mathbb{Z}}L^k e^{-\frac{k^2\beta^2}{4s}}\\ &=& (4\pi\mu^2)^\epsilon \bigg[\Gamma(\epsilon-2)m^{2(2-\epsilon)} \nonumber\\
&+& 2\sum_{k=1}^{\infty} L^k\left(\frac{k^2\beta^2}{4\widetilde{m}^2}\right)^{-\frac{2-\epsilon}{2}} K_{-2+\epsilon}(\widetilde{m}k\beta)\bigg],\label{intgtnresult}
\end{eqnarray}
where we have used the formula~\cite{gradshteyn} 
\begin{eqnarray}
\int_0^{\infty} dx~ x^n e^{-\lambda x}= \frac{n!}{\lambda^{n+1}}, \quad [\text{Re}~\lambda>0],
\end{eqnarray}
and the  modified Bessel function of ~$K_n(x)$ in Eq.~(\ref{intgtnresult}) appears through its integral representation:
 \begin{eqnarray}
 \int_0^\infty x^{n-1} e^{-\frac{\beta}{x}-\alpha x} dx= 2\left(\frac{\beta}{\alpha}\right)^{\frac{n}{2}}~K_n(2\sqrt{\beta\alpha}), \nonumber\\ \quad [\text{Re}~\alpha>0,~ \text{Re}~\beta >0].
 \end{eqnarray}
 The first term in Eq.~(\ref{intgtnresult}) appears in any massive field theory at zero temperature. But, the next term is interesting because of its dependence on the untraced Polyakov loop (L) at finite temperature. The $L$-dependence of the EMT at high temperature will be discussed later. It is one of the main results of our present investigation. For the effective action at finite temperature, in the limit $\epsilon\to 0 $, $\beta\to 0$,  we have
 \begin{eqnarray}
 W^{1(T\neq 0)}_0 = \int d^4x \sqrt{-\widetilde{g}}~ \bigg[- \frac{\pi^2}{45}T^4(N^2-1) \nonumber\\
 + \,\mathcal{O}(L, \widetilde{m}\beta)\bigg],
 \label{effectiveactnTMM}                      
 \end{eqnarray}
where the leading order term is matched with the expected result in Eq.~(\ref{modeffectiveactn}). In getting the above  expression, we have used $K_{-n}(x)= K_n(x)$~\cite{gradshteyn}. In the above, $\mathcal{O}(L, \widetilde{m}\beta)$ designates the terms which contain various  non-zero power of $L$ and dimensionless quantity $\widetilde{m}\beta$ 
(with $\beta=T^{-1}$). The appearance of leading order terms can be understood from the behaviour of $K_n(x)$ for small argument $(x\to 0)$ as~\cite{gradshteyn}
\begin{eqnarray}
K_n(x)\sim \frac{1}{2}\Gamma(n)\left(\frac{x}{2}\right)^{-n},
\end{eqnarray}
and the expansion of the time-ordered product is
\begin{eqnarray}
L^k(x, \beta)&=& \left(1-\int_{x_0}^{x_0+\beta} a_0(x'_0, x)L(x',\beta) dx'_0\right)^k \nonumber\\
&=&1-k\int_{x_0}^{x_0+\beta} a_0(x'_0, \mathbf{x})L(x', \beta)~dx'_0+\cdots, \quad
\end{eqnarray}
where $x'\equiv (x'_0, \mathbf{x})$.

We also obtain  another important result which carries a great significance in a  renormalizable  massive non-Abelian gauge theory. From the dimensional regularization for the coefficient $a_2^{T=0}$, we obtain
\begin{eqnarray}
W^{1(T=0)}_2=\lim_{\epsilon\to 0}  g^2 N\frac{1}{(4\pi)^2}\int d^4x \sqrt{-\widetilde{g}} \frac{1}{\bar{\epsilon}} \frac{11}{12} F^{a\mu\nu} F^a_{\mu\nu},
\end{eqnarray} 
where $\dfrac{1}{\bar{\epsilon}}=\left(\dfrac{1}{\epsilon}-\gamma_E+\ln4\pi\right)$ and $\gamma_E\approx 0.5772$ is called to be Euler's constant. This result appears from the contribution of $a_2^{T=0}$. This contribution is same as found in the massless YM theory. Thus, we  get the same asymptotic behaviour (i.e. asymptotic freedom) of gauge coupling  $g$ in the non-Abelian TMM. Therefore, the applicability of the perturbative technique at high temperature is consistent.

\begin{figure}[h]
\centering
\includegraphics[width=8.9cm]{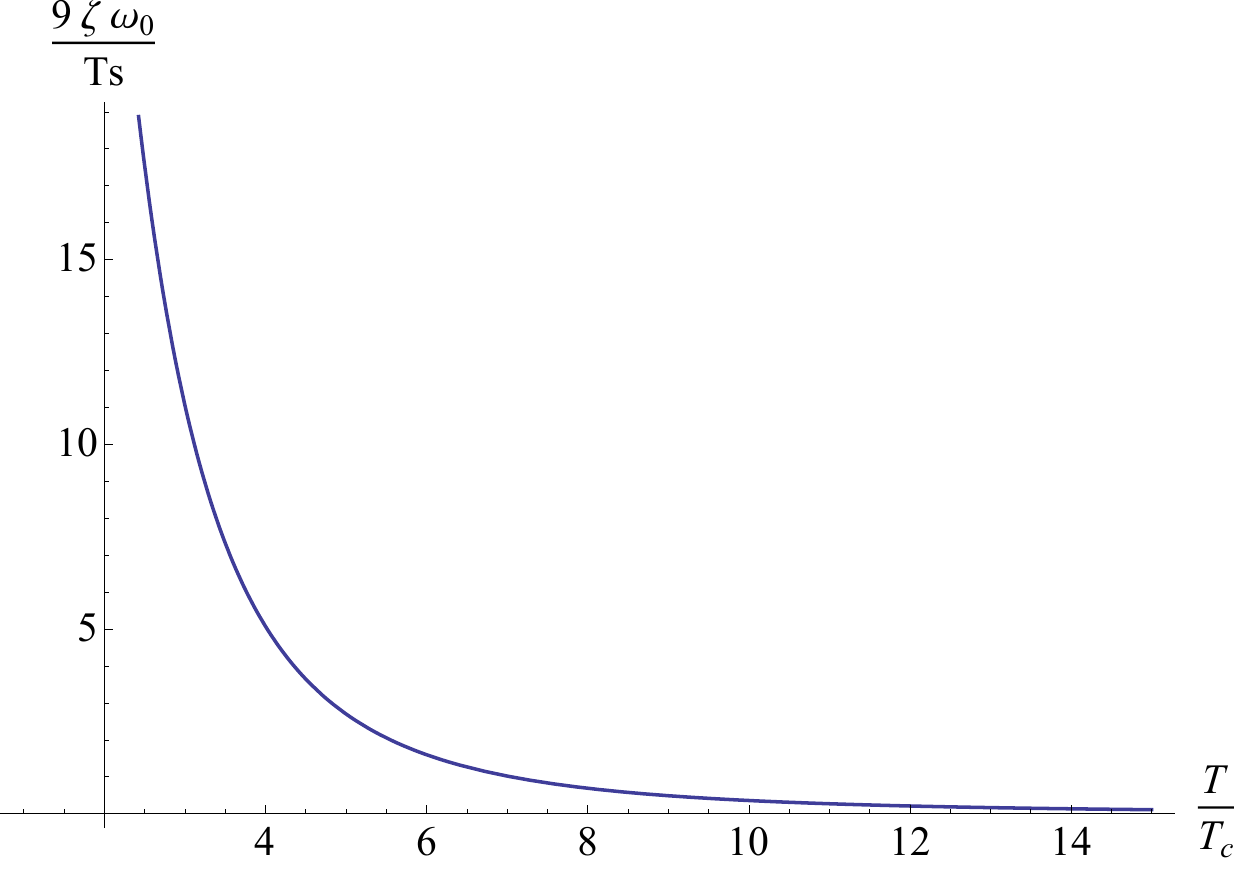}
\caption{The variation of $\frac{9\zeta\omega_0}{sT}$ with $T/T_c$ with $\omega_0=1.5 {\text GeV}$. 
}
\label{fig1}
\end{figure}

Now we go back to the low energy theorem at finite temperature [cf. Eq.~(\ref{let})]. For this purpose, we consider  Eq.~(\ref{let}) for $n=1$:
\begin{eqnarray}
\left(T\frac{\partial}{\partial T}-4\right)\braket{\Theta^\mu_\mu}
&=& \int d\tau d^3x \braket{\Theta^\mu_\mu(\tau, x)\Theta^\mu_\mu(0, 0)} \nonumber\\
&&+\, \frac{(1-3c_s^2)}{c_s^2}h,
\label{keyeqn}
\end{eqnarray}
 where, we add a term in the right hand side of the above equation. This term is appeared from the consideration of right hydrodynamic limit to get the transport coefficients ~\cite{romatschke}. $h$ is enthalpy density and $c_s$, the speed of sound. Now, putting the expression of $a^T_0$ [cf. Eq.~(\ref{htkrnlceff0})] in the above expression, we get in the high temperature limit
\begin{widetext}
\begin{eqnarray}
\int d\tau d^3x \braket{\Theta^\mu_\mu(\tau, x)\Theta^\nu_\nu(0, 0)} &=&-\frac{g^2(N^2-1)}{(4\pi)^2}\sum_{n=1}^\infty\frac{4\widetilde{m}^2}{n^2}\left[-\frac{n\widetilde{m}}{\beta}K_1(n\widetilde{m}\beta) \Braket{\text{tr}~L^n}+\frac{2}{n \widetilde{m}^2\beta^3}\Braket{\text{tr}\left(L'L^{n-1}\right)}+\cdots\right]\nonumber \\ 
&=&-\frac{g^2(N^2-1)}{(4\pi)^2}\sum_{n=1}^\infty \frac{4T^2\widetilde{m}^2}{n^2}\left[\frac{2T}{n\widetilde{m}^2}\braket{\text{tr}\left(L' L^{n-1}\right)}-\braket{\text{tr}~L^n}+\cdots\right],
\label{emtcorreltn}
\end{eqnarray} 
where $L'=\dfrac{\partial L}{\partial\beta}$. In the last step of the above equation, we have used a series expansion of the modified Bessel function~\cite{gradshteyn}:
\begin{eqnarray}
K_n(z)&=&\frac{1}{2}\left(\frac{1}{2}z\right)^{-n}\sum_{k=0}^{n-1}\frac{(n-k-1)!}{k!}\left(-\frac{z^2}{4}\right)^k-(-1)^{n}\ln\left(\frac{1}{2}z\right)I_n(z) \nonumber\\
&+& (-1)^n\frac{1}{2}\left(\frac{z}{2}\right)^n\sum_{k=0}^{\infty}\{\psi(k+1)+\psi(k+n+1)\}\frac{\left(\frac{z^2}{4}\right)^k}{k!(n+k)!},
\label{seriesexpntn}
\end{eqnarray}
\end{widetext}
and the following recursion relation~\cite{gradshteyn} has also been utilized:
\begin{eqnarray}
\frac{d K_n(z)}{dz}= -K_{n-1}-\frac{n}{z} K_n(z).
\end{eqnarray}
In the series expansion [cf. Eq.~(\ref{seriesexpntn})], $I_n(x)$ is the modified Bessel function ~\cite{gradshteyn}:
\begin{eqnarray}
I_n(x)=\left(\frac{x}{2}\right)^n\sum_{k=0}^{\infty}\frac{\left(\frac{x^2}{2}\right)^k}{k!\Gamma(n+k+1)}
\end{eqnarray}
and $\psi(n)$ is Euler's $\psi$ function or diagmma function defined as~\cite{gradshteyn} 
\begin{eqnarray}
\psi(x)=-\gamma_E-\sum_{k=0}^{\infty}\left(\frac{1}{x+k}-\frac{1}{k+1}\right).
\end{eqnarray}
Introducing the spectral function $\rho(\omega)$ as 
\begin{eqnarray}
\int d\tau d^3x \braket{\Theta^\mu_\mu(\tau, x)\Theta^\nu_\nu(0, 0)}= 2\int_0^{\infty}\frac{\rho(\omega, \vec{0})}{\omega}~d\omega,
\label{spectral}
\end{eqnarray}
and taking the expression of $\rho(\omega, \vec{0})$  for small frequencies~\cite{karsch}
\begin{eqnarray}
\frac{\rho(\omega, \vec{0})}{\omega}= \frac{9\zeta_T}{\pi}\frac{\omega_0^2}{\omega^2+\omega_0^2},
\end{eqnarray}
we get\footnote{Retarded correlation between the traces of energy momentum tensors cannot be distinguished from the correlation involving commutator of the traces in linear response theory (see~\cite{jeon} for details).} from Eq.~(\ref{emtcorreltn})
\begin{eqnarray}
9\omega_0 \zeta_T(\omega_0)&=-&\frac{g^2(N^2-1)}{(4\pi)^2}  \sum_{n=1}^\infty \frac{8T^3}{n^2}\bigg[\Braket{\text{tr}\left (L' L^{n-1}\right)}\nonumber\\ 
&& -\,\frac{\widetilde{m}^2}{2T}\Braket{\text{tr}~L^n}+\cdots\bigg] + \frac{(1-3c_s^2)}{c_s^2}h.
\label{bulkviscosity}
\end{eqnarray}
The above equation shows the dependence of $\zeta_T$ on the mass of the gluon and thermal average 
value of various power of untraced Polyakov loop in the deconfined phase. 
The bulk viscosity can be calculated by using Eq.~\ref{bulkviscosity} with known expression 
for the Polyakov loop. For simplicity, we use the analytical expression for the Polyakov 
loop of pure $SU(3)$ field obtained by using gauge-string duality in~\cite{OAndreev} 
to estimate  $\zeta_T$. The expression for Polyakov loop given in Ref.~\cite{OAndreev} 
reproduces the lattice QCD results reasonably well. 
In Fig.~\ref{fig1} the variation of the ratio, 
$\dfrac{9\zeta\omega_0}{sT}$ with  $\dfrac{T}{T_c}$ is displayed.
The nature of variation is similar to that obtained in Ref.~\cite{karsch}.
Here $T_c$ is the critical temperature for quark-hadron transition and
$s$ is the entropy density which is estimated as follows. 
We have already observed from the expression of effective action
in Eq.~\ref{effectiveactnTMM} that 'effectively' the transverse 
degrees of freedom (tDOF)  
of gluons participate in the leading term due to the combined contributions
of ghost sectors corresponding to the one-form ($A_\mu$) and two-form 
($B_{\mu\nu}$) fields (Refs.~\cite{dewitt}, \cite{duff}, \cite{siegel}). 
Consequently, the entropy of gluonic fluid
is also constituted by the same contributions from the two tDOFs and
eight colour degrees of freedom, that is, 
$s=4\frac{\pi^2}{90}gT^3$, $g=2\times 8$ is the statistical degeneracy 
of the gluons.

The results obtained in this work can be applied to a system of
pure gluonic matter only i.e. to a system which can be described by
non-zero temperature and zero baryonic chemical potential ($\mu_B$). 
However, there are outstanding physics issues to be addressed 
for thermal QCD system at non-zero $T$ and $\mu_B$.   One such issue 
is the existence and detection of the  
critical point in the QCD phase diagram at non-zero $T$ and $\mu_B$
~\cite{Fodor2004}.  The extension of the current formalism to 
the domain of non-zero chemical potential will help us to 
understand the behaviour of bulk viscosity near the
critical point.  In such case 
the variation of $\braket{\text{tr}~L}$, which is considered as an ordered parameter for 
confinement to deconfinement transition, with temperature and baryonic chemical potential 
will also govern the variation of $\zeta_T$ near the transition point.
Therefore, this relation will be useful to understand the variation of bulk viscosity with 
temperature and chemical potential to determine the value of critical exponent of bulk viscosity 
at the critical point of QCD-phase transition.

Commonly, the following  procedure is used for estimating the bulk viscosity
near the critical point. The  LRT is used to calculate it away from the
critical point. 
The bulk viscosity  near the critical point ($\zeta_T^{\text cr}(T,\mu_B)$)
is obtained  then by using the following scaling behaviour
($\xi$)~\cite{Monnai2017}:
\begin{equation}
\zeta_T^{\text cr} = \zeta_T \left( \frac{\xi}{\xi_0}\right)^3 
\label{zeta_cr}
\end{equation}
where $\zeta_T$ is the bulk viscosity away from the
critical point, $\xi(T,\mu_B)$ is the correlation length 
which diverges near the critical point and $\xi_0$ is a constant, 
typically, $\xi_0\sim 1.75$ fm~\cite{Monnai2017}.  

It is worth mentioning that within the scope of the formalism adopted
in the present work to estimate bulk viscosity, the effects of the critical point
will infiltrate to bulk viscosity through the behaviour of the  Polyakov loop near
the critical point. 

\section{Summary and Discussion} 
\label{Section3}
We have found the bulk viscosity  $\zeta_T$ within the scope of TMM and its dependence on the thermally averaged untraced Polyakov loop as well as its various powers and derivative analytically in Eq.~(\ref{bulkviscosity}). We observe that $\zeta_T$ is positive in every order of quantum corrections. It is because of the BRST invariance of effective action in the quantum 
corrections which is a consequence of the renormalizibility of the model. This causes the maintenance of the convexity of effective potential~\cite{sw} in the corrections. The positivity of $\zeta_T$ is required to obey the second law of thermodynamics~\cite{weinberg}. The dependence on the various power of the thermal expectation value of $L$  [cf. Eq.~(\ref{bulkviscosity})] appears from the LET where correlation among trace of energy momentum tensor densities is involved. In Eq.~(\ref{bulkviscosity}),  the terms containing $\Braket{\text{tr}~L^n}$ and $\Braket{\text{tr}~\left(L' L^{n-1}\right)}$ are not invariant under $\mathbb{Z}_N$ group\footnote{The elements of the $\mathbb{Z}_N$ is $z=e^{\frac{2\pi i n}{N}}\mathbf{1}$, where $n=0, 1, 2, \ldots, (N-1)$; $\mathbf{1}$ designates a $N\times N$ unit matrix.}, which is the centre of $SU(N)$ group. As a consequence,  the contribution of $\zeta_T$ will also be significant in the study of QGP at heavy  quark limit where the restoration of $\mathbb{Z}_N$ symmetry implies the phase transition of QGP i.e. deconfined phase to confined phase.

Here we should make comments from our observations on a puzzle raised in~\cite{moor2}. The authors in ~\cite{moor2} pointed out a mismatch of the power of gauge coupling  in the sum rule  that is given in~\cite{karsch}. This issue was addressed in~\cite{hout} by considering an operator mixing in renormlization group approach. Generally, an operator product expansion is made in the deep ultraviolet region of Euclidean momentum space. The ultraviolet behaviour of various Green functions or correlators depend on their off-shell behaviour. But in thermal field theory, the real and imaginary time formalisms show that the off-shell nature of correlators, which causes the renormalization of the fields and couplings, is independent of 
temperature~\cite{das,laine}. Hence, the Callan-Symanzik renormalization group equation is always satisfied in a renormalizable gauge theory.  For example, the $n$-point Green functions after quantum corrections generally takes the form in Lorentz covariant way~\cite{mallik}:  
\begin{eqnarray}
\Gamma_R^{\mu\nu\ldots}(x_1, x_2,\ldots, x_n,T )=\Gamma_R^{\mu\nu\ldots}(x_1, x_2,\ldots, x_n, 0 )\nonumber\\
+u^\mu u^\nu\ldots \Delta^i\Gamma((x_1, x_2,\ldots, x_n,T )+\cdots,
\end{eqnarray}
where the subscript $R$ designates the renormalized n-point function, $\Delta^i\Gamma$ represents the $i$-th order correction and $u^\mu$ is the four velocity of heat bath. It is interesting to note  that the real and imaginary time formalisms in TFT provide
the inequivalent $3$-point functions~\cite{evans, kobes}. The quantum correction of 3-point vertex in pure YM theory at finite temperature in real time formalism leads us, logically, to the dependence of the gauge coupling $g$ on temperature~\cite{fujimoto, bair, ashida}. This dependence shows that $T\dfrac{d\beta}{dT}$ is not proportional to $g^6$ at the leading order even in the case of massless YM theory. Rather, $T\dfrac{d\beta}{dT}\propto g^4$ at leading order. The puzzle will also never arises in the model that is considered in our the present work. It is because, the model does not provide any trace anomaly.  Hence, at the leading order, the both side of the Eq.~(\ref{keyeqn}) [cf. Eq.~(\ref{spectral})] are proportional to $g^2$.

The authors in~\cite{romatschke} have pointed out the domain of validity of the low energy theorem~\cite{karsch} which is used in this work. 
However, the conformally Minskowski flat  metric used in                                                                                                                                                    ~\cite{romatschke} is not consistent with the phenomenology of general relativity in small scale\footnote{Here `small scale' implies the scale which is much less than the cosmological scale.}~\cite{hawking}. Besides this, there is no ``physically meaningful'' unique renormalized EMT in curved spacetime~\cite{wald78} for massless 
fields~\cite{amir,kay}. It is because the required Hadamard state has non-local singularity for massless field and there is no unique de-Sitter group invariant vacuum state of this field~\cite{allen}. 

Now, we are going to discuss on the results where we have reached in the last section. In arriving at Eq.~(\ref{bulkviscosity}), we have obtained two very significant results for QCD: 
\begin{enumerate}[label=(\roman*)]    
\item The leading terms in the expression of effective action [cf. Eq.~(\ref{effectiveactnTMM})] 
match  with the leading terms for the massless YM theory [cf. Eq.~(\ref{effactnmasslessYM})]. This equality is due to the resultant null contribution from the kinetic terms of $B$, $\omega_\mu$ and $\bar{\omega}_\mu$, $\beta$ and $\bar{\beta}$, and $\widetilde{n}$ (i.e. $B$ field sector)[cf. Eq.~(\ref{nonabactn})]. It can be understood by counting the total degrees of freedom of the fields~\cite{duff, siegel}  in the $B$ field sector, contributing in the effective action in 3+1 dimensions: $1\times 6- 2\times 4+2\times 1=0$. This resultant null contribution occurs because the kinetic term of $\widetilde{n}$ does not contain any covariant derivative [cf. Eq.~(\ref{nonabactn})].

\item The asymptotic freedom remains same as found in massless YM field theory. It (with the IR cut-off) assures the validity of the calculation of bulk viscosity in the perturbative regime at non-zero temperature. We also note that the resuumation~\cite{kapusta} are absent due to the presence of IR cut-off in the TMM. As a consequence, the terms originated with odd power of $g$ or fractional power of strong coupling $\alpha_s =\frac{g^2}{4\pi}$, like $\mathcal{O}\left(g^3\right)\sim\mathcal{O}\left(\alpha_s^{\frac{3}{2}}\right)$ and the terms involving $\alpha_s^2 \ln \alpha_s$,  etc., in the expression of pressure in massless YM theory at high temperature~\cite{kapusta}, is absent in the case of TMM. The appearance of those terms in the analysis ensures the breakdown of analytic property of perturbative theory according to~\cite{dyson, bellac}. The absence of the terms$\sim\mathcal{O}\left(\alpha_s^{\frac{3}{2}}\right)$ in the effective action shows that the particle number changing process is slower than the massless YM theories~\cite{moor1}.

\end{enumerate}

The present result has been obtained in the realm of perturbative approach. Even though, the hadronization is a non-perturbative process, we can make the following concluding remark. The variation of bulk viscosity near the transition point is governed by Polyakov loop. The enhancement of bulk viscosity near the transition point will reduce the effective pressure of the fluid which, in turn, will provide a smaller kick (as opposed the case when $\zeta_T=0$) to the produced particles. This would be reflected in the experimentally measured value of average transverse momentum of the hadrons. Moreover, the reduced pressure will slow down the expansion resulting in production of more soft gluons enhancing the multiplicity of produced hadrons. Therefore, the present work indicates a possibility to measure Polyakov loop experimentally. It is worth mentioning here that calculations, based on the lattice QCD ~\cite{okaczmarek,sgupta}, indicate a sharp decrease of 
Polyakov loop with temperature near critical point. This causes a sharp rise of $-\langle \text{tr}\, L^\prime\rangle=-\bigg\langle \text{tr} \,\dfrac{\partial L}{\partial\beta} \bigg\rangle$
appeared in Eq.~(\ref{bulkviscosity}). Hence, it implies a large increase of bulk viscosity near the critical point which is expected in phase transition.  Such variation of bulk viscosity   
is consistent with the results obtained 
from calculations based on lattice QCD ~\cite{astrakhantsev}
and phenomenological model~\cite{nicola}.

The present investigation may play an important role in the study of early universe and its evolution. In the M\"{u}ller-Israel-Stewert theory of causal hydrodynamics~\cite{muller,israel1, israel2, paul}, it will be interesting to observe the importance of broken $\mathbb{Z}_N$ symmetry through the dependence of entropy production rate on $\zeta_T$ at the time of QGP phase transition. We can also note that the contribution of $\zeta_T$ from TMM  will be different from the case of massless YM theory in entropy production rate due to absence of  resummation. Other transport coefficients are remained to be calculated from the TMM at finite temperature, whose behaviours at large-$N$ limit can be investigated. The significance of Eq.~(\ref{bulkviscosity}) can also be explored in the scenarios of bulk viscous cosmology~\cite{zim1, zim2} for the study of dark matter and dark energy. 


\noindent
\section*{Acknowledgments}
DM is thankful to the Department of Atomic Energy, Government of India for financial support. 
RK would like to thank the University Grants Commission, Government of India, New Delhi, for financial support
under the PDFSS scheme.




\begin{thebibliography}{99}
\bibitem{momen} A. Momen, 
Phys. Lett. B {\bf 394},  269 (1997).

\bibitem{mann}  R.B. Mann, 
JHEP {\bf 06},  075 (2009).


\bibitem{jerzy} J. Kowalski-Glikman, 
Phys. Rev. D {\bf 81}, 084038 (2010).


\bibitem{cs:1974} E. Cremmer and J. Scherk, 
Nucl. Phys. B {\bf 72}, 117 (1974).




\bibitem{aab:1990} T.J. Allen, M.J. Bowick and A. Lahiri, 
Mod. Phys. Lett. A {\bf 6}  559 (1990).


\bibitem{meig:1983} L. Baulieu and J. Thierry-Mieg,
Nucl. Phys. B {\bf 197}, 477 (1982). 



\bibitem{al:1997}  A. Lahiri,
Phys. Rev.  D {\bf 55},  5045 (1997).


\bibitem{al:01} A. Lahiri,
Phys. Rev. D  {\bf 63},   105002 (2001).


\bibitem{kr:1973}  M. Kalb and  P. Ramond,
Phys. Rev. D {\bf 9},   2273 (1973).

\bibitem{dewitt} Bryce S. DeWitt, {\it The global approach to quantum field theory}, Vols. 1 and 2, 
(Oxford University Press, 2003).

\bibitem{rohit1} R. Kumar and R. P. Malik, 
Eur. Phys. J. C {\bf 71}, 1710 (2011).


\bibitem{rohit2} R. Kumar and D. Mukhopadhyay,
Euro Phys. J C {\bf 78}, 452 (2018).


\bibitem{stro} F. Strocci, 
Phys. Lett. B {\bf 62},  60 (1976).

\bibitem{haag} R. Hagg and  D. Kastler, 
J. Math. Phys. {\bf 5},  848 (1964).




\bibitem{kugo}  T. Kugo and  I. Ojima, 
Prog. Theor. Phys. Suppl. {\bf 66},  1 (1979).



\bibitem{fischer} C.S. Fischer, 
J. Phys. G: Nucl. Part. Phys.  {\bf 32}, R253 (2006).

\bibitem{chaichian} M. Chaichian and K. Nishijima,  
Eur. Phys. J.  C {\bf 47},   737 (2006).




\bibitem{gross}  D.J. Gross and  F. Wilczek, 
Phys. Rev. Lett. {\bf 30},  1343 (1973).

 
\bibitem{politzer}  H.D. Politzer, 
Phys. Rev. Lett. {\bf 30}, 1346 (1973).  


\bibitem{sc:1973}  S. Coleman and  D.J. Gross, 
Phys. Rev. Lett. {\bf 31},   851 (1973).




\bibitem{linde} A.D. Linde, 
Phys. Lett. B {\bf 96}, 289 (1980). 


\bibitem{furusawa} T. Furusawa and K. Kikkawa, 
Phys. Lett. B {\bf 128},  218 (1983). 

\bibitem{kapusta} J.I. Kapusta, {\it Finite temperature field theory} (Cambridge University Press, USA, 1993).


\bibitem{bellac} M. Le Ballac, {\it Thermal field theory} (Cambridge University Press, USA, 1996).


\bibitem{billoire}
 A. Billoire, G. Lazarides and Q. Shafi, 
 Phys. Lett. B {\bf 103}, 450 (1981).
 
 \bibitem{degrand}
 T.A. Degrand and D. Toussaint
 Phys. Rev. D {\bf 25},  526 (1982).

\bibitem{fukuda} R. Fukuda, 
Phys. Lett. B {\bf 73},  33 (1978). 

\bibitem{huang} K. Huang, Quarks, {\it leptons and gauge fields}  (World Scientific,  1992).


\bibitem{debmalya} D. Mukhopadhyay, R. Kumar, J. Alam, and S.K. Singh, Phys. Rev. D {\bf 101}, 074039 (2020).

\bibitem{Busza} W. Busza, K. Rajagopal and W. van der Schee
Ann. Rev. Nucl. Part. Sci. {\bf 68}, 339 (2018). 

\bibitem{CBM} B. Friman, C. H\"ohne, J. Knoll, S. Leupold, 
J. Randrup, R. Rapp and P. Senger (Eds.),
The CBM Physics Book: Compressed Baryonic Matter in 
Laboratory Experiments, Springer, 2011.

\bibitem{jeon} S. Jeon, 
Phys. Rev.  {\bf 52}, 3591 (1996).

\bibitem{moor1} P. Arnold, C. Dogan and G.D. Moore, 
Phys. Rev. D {\bf 74}, 085021  (2006).

\bibitem{moor2} G.D. Moore and  O. Saremi,  
JHEP {\bf 0809},  015 (2008).



\bibitem{bernhard}
J.E. Bernhard, J.S. Moreland and S.A. Bass, 
Nature Phys. {\bf 15},  1113 (2019).

\bibitem{alford}
M. Alford, A. Harutyunyan and A. Sedrakian, Phys. Rev. D {\bf 100}, 103021 (2019).

\bibitem{bemfica}
F.S. Bemfica, M.M. Disconzi and J. Noronha, Phys. Rev. Lett. {\bf 122},  221602 (2019).

\bibitem{czajka}
A. Czajka , K. Dasgupta, C. Gale, S. Jeon, A. Misra , M. Richard and K. Sil, JHEP {\bf 1907},  145 (2019).

\bibitem{astrakhantsev}
N.Yu. Astrakhantsev, V.V. Braguta and A.Yu. Kotov,  Phys. Rev. D {\bf 98},  054515 (2018).

\bibitem{hattori}
K. Hattori, Xu-Guang Huang, D.H. Rischke and D. Satow, Phys. Rev. D {\bf 96},  094009  (2017).


\bibitem{Ryu} S. Ryu, J.-F. Paquet, C. Shen, G.S. Denicol, B. Schenke, S. Jeon and C. Gale,
Phys. Rev. Lett. {\bf 115},  132301 (2015).


\bibitem{meyer}
H.B. Meyer,
Phys. Rev. Lett. {\bf 100},  162001 (2008).

\bibitem{bala1} V. Balakrishnan, 
{\it Elements of nonequilibrium statistical mechanics}  (CRC Press, US, 2008).


 \bibitem{benincasa}  P. Benincasa and A. Buchel, 
JHEP {\bf 0601},  103 (2006).


\bibitem{jeon1} S. Jeon and U. Heinz,
Int. Mod. Phys. E 24 (2015) 1530010.

\bibitem{bala2} V. Balakrishnan, {\it Mathematical physics with applications, problems and solutions} (Ane Books, UK, 2019).




\bibitem{dowker} J.S. Dowker and R. Critchley, 
Phys. Rev. D {\bf 16}  3390 (1977).

\bibitem{wald} R.M. Wald, 
Commun. Math. Phys. {\bf 54}, 1 (1977). 


\bibitem{stelle} K.S. Stelle,  
Gen. Rel. Grav.  {\bf 9}, 535 (1978). 


\bibitem{shifman}
M. A. Shifman, {\it Vacuum structure and QCD sum rules} (Elsevier, Amsterdam, 1992).

\bibitem{ralf}
R. Sch\"utzhold, Phys. Rev. Lett. {\bf 89}, 081302 (2002).



\bibitem{sebastein} J.-S.  Gagnon and  J. Lesgourgues, 
 JCAP {\bf 09},  026 (2011).



\bibitem{ellis} P.J. Ellis, J.I. Kapusta and H.-B. Tang, 
Phys. Lett.  B {\bf 443},  63 (1998).

\bibitem{sushpanov} I.A. Sushpanov, J.I. Kapusta and P.J. Ellis, 
Phys. Rev. C {\bf 59},  2931 (1998).

\bibitem{karsch} F. Karsch, D. Kharzeev and K. Tuchin,
Phys. Lett. B {\bf 663},  217 (2008). 


\bibitem{christensen1} S.M. Christensen, 
Phys. Rev. D {\bf 14},  2490  (1976).

\bibitem{christensen2} S.M. Christensen, 
Phys, Rev. D {\bf 17}, 946 (1978).





\bibitem{barvinsky} A.O. Barvinsky and G.A. Vilkovisky, 
Phys. Rep. {\bf 119}, 1 (1985). 


\bibitem{filho} H. Boschi-Filho,  C.P. Natividade and C. Farina, 
Phys. Rev. D {\bf 45}, 586 (1992).


\bibitem{Vassilevich} D.V. Vassilevich,
Phys. Rep.  {\bf 388},  279 (2003). 

 
\bibitem{megias1} E. Megias, E.R. Arriola and L.L. Salcedo,
Phys. Lett.  B {\bf 563},  173 (2003).


\bibitem{megias2} E. Megias, E.R. Arriola and  L.L. Salcedo, 
Phys. Rev.  D {\bf 69},   116003 (2004).

\bibitem{polyakov}  A.M. Polyakov, 
Phys. Lett.  B {\bf 72},  477 (1978).
 



\bibitem{sw} S. Weinberg, {\it The quantum theory of fields}, Vol. 2 (Cambridge University Press, NY, 1996).

\bibitem{gradshteyn} I.S. Gradshteyn and I.M. Ryzhik, {\it Tables of integrals and products} 
(Academic Press, London, 1963).


\bibitem{romatschke}
P. Romatschke and D.T. Son, Phys. Rev. D {\bf 80}, 065021 (2009).

\bibitem{OAndreev} O. Andreev, Phys. Rev. Lett. {\bf 102}, 212001 (2009).

\bibitem{Fodor2004} Z.~Fodor and S.~Katz, JHEP \textbf{04}, 050 (2004).

\bibitem{Monnai2017} A. Monnai, S. Mukherjee and Y. Yin, Phys. Rev. C {\bf 95}, 034902 (2017).


\bibitem{weinberg} S. Weinberg, 
Astrophys. J.  {\bf 168}, 175 (1971).

\bibitem{hout}
S. Coron-Hout. Phys. Rev. D {\bf 79}, 125009 (2009). 


\bibitem{das} A. Das, {\it Finite Tempretaure Field Theory} (World Scientific, Singapore, 1997).

\bibitem{laine}
M. Laine and A. Vuorinen, {\it Basics of Thermal Field Theory}
(Springer, Switzerland, 2016).

\bibitem{mallik}
S. Mallik, Phys. Lett. B {\bf 416}, 373 (1998). 

\bibitem{evans}
T.S. Evans, Phys. Lett. B {\bf 249}, 286 (1990).

\bibitem{kobes}
R. Kobes, Phys. Rev. Lett. {\bf 67}, 1384 (1991). 


\bibitem{fujimoto}
Y. Fujimoto and H. Yamada, Phys. Lett. B {\bf 195}, 231 (1987).

\bibitem{bair}
R. Bair, B. Pire and D. Schiff, Phys. Rev. D {\bf 38}, 2566 (1990).

\bibitem{ashida}
N. Ashida, A, Niegawa,  H. Nakkagawa and H. Yokota, Phys. Rev. D {\bf 44}, 473 (1991).


\bibitem{hawking}
S.W. Hawking and G.F.R. Ellis, {\it The large scale structure of space-time} (Cambridge Univ. Press, USA, 1994).

\bibitem{wald78}
R.M. Wald, Phys. Rev. D {\bf 17}, 1477 (1978). 

\bibitem{amir}
A.H. Najmi and A. C. Ottewill, Phys. Rev. D {\bf 32}, 1942 (1985).

\bibitem{kay}
G. Gonnella and B. S. Kay, Class. Quant. Grav. {\bf 6}, 1445 (1989)

\bibitem{allen}
B. Allen, Phys. Rev. D {\bf 32}, 3136 (1985).

\bibitem{duff} M.J. Duff and P. van Nieuwenuizen, 
Phys. Lett.  B {\bf 94},  179 (1980).


\bibitem{siegel} W. Siegel, 
Phys. Lett. B {\bf 103},   107 (1981). 

\bibitem{dyson}
F. D. Dyson, Phys. Rev. {\bf 85}, 631 (1951).

\bibitem{okaczmarek} O. Kaczmarek, F. Karsch, P. Petreczky, and F. Zantow, Phys. Lett. B {\bf 543}, 
41 (2002). 

\bibitem{sgupta} S. Gupta, K. Huebner, and O. Kaczmarek, Phys. Rev.
D {\bf 77},  034503 (2008).



\bibitem{nicola} D. Ferna\'ndez-Fraile and A. G. Nicola, Phys. Rev. Lett. {\bf 102}, 121601 (2009).


\bibitem{muller} I. M\"{u}ller, 
Z. Phys. {\bf 198},  329 (1967).

\bibitem{israel1} W. Israel, 
Ann. Phys. {\bf 100}, 310 (1976).

\bibitem{israel2} W. Israel and J.M. Stewart, 
Ann. Phys. {\bf 118},  341 (1979).

\bibitem{paul} P. Romatschke, 
Int. J. Mod. Phys. E {\bf 19}, 1 (2010).


\bibitem{zim1} W. Zimdahl, 
Phys. Rev. D {\bf 53}, 5483 (1995).


\bibitem{zim2} O.F. Piattella, J.C. Fabries and W. Zimdahl, 
JCAP {\bf 05},   029 (2011).




\end{thebibliography}
\end{document}